Validation of density functionals for transition metals and intermetallics using data from quantitative electron diffraction

Xiahan Sang, Andreas Kulovits, Guofeng Wang, Jörg Wiezorek

*Department of Mechanical Engineering and Materials Science, University of Pittsburgh, Pittsburgh, Pennsylvania 15261*

Abstract

Accurate low-order structure factors ($F_g$) measured by quantitative convergent beam electron diffraction (QCBED) were used for validation of different density functional theory (DFT) approximations. 23 low-order $F_g$ were measured by QCBED for the transition metals Cr, Fe, Co, Ni, and Cu, and the transition metal based intermetallic phases γ-TiAl, β-NiAl and $γ_1$-FePd using a multi-beam off-zone axis (MBOZA) method and then compared with $F_g$ calculated *ab-initio* by DFT using the local spin density approximation (LDA) and LDA+U, and different generalized gradient approximations (GGA) functionals. Different functionals perform very differently for different materials and crystal structures. Among the GGA functionals, PW91 and EV93 achieve the best overall agreement with the experimentally determined low-order $F_g$ for the five metals, while PW91 performs the best for the three intermetallics. The LDA+U approach, through careful selection of U, achieves excellent matches with the experimentally measured $F_g$ for all the metallic systems investigated in this paper. Similar to the band gap for semiconductors, it is proposed that experimentally determined low-order $F_g$ can be used to tune the U term in LDA+U method DFT calculations for metals and intermetallics.

PACS number(s): 71.15.Mb, 71.20.Be, 71.20.Lp, 61.05.jm

## I. INTRODUCTION

Density functional theory (DFT) calculation has been widely used in the field of computational materials science due to its accuracy and relatively affordable requirements regarding computer time.[1,2] However, an intricate part in DFT, the exact formula of the exchange-correlation energy, $E_{xc}$, remains unknown and can only be approximated with some functionals of electron density $n(\mathbf{r})$. For instance, $E_{xc}$ can be approximated by a functional of $n(\mathbf{r})$ as in the local density approximation (LDA).[2] In contrast, generalized gradient approximation (GGA) functionals include a gradient of $n(\mathbf{r})$ and thereby improve on LDA.[3] Meta-GGA functionals further include the Laplacian of $n(\mathbf{r})$.[4, 5] Improvements over LDA have also been achieved by the inclusion of orbital dependent potential terms, for example LDA+U,[6, 7] which exhibit significant improvements in the prediction of band gaps of semiconductors when compared to conventional LDA calculations.[8] Various DFT functionals have been specifically designed for different materials systems and applications to achieve optimized performance regarding predictions of a property of interest, e. g. the electronic band gap in semiconductors[9, 10] or lattice constants.[11] While the functionals thus obtained might therefore be able to calculate accurate values for the properties they are designed to predict, additional independent validation is essential before they can be used with confidence for other applications and purposes. Currently, most validation approaches focus on comparison of properties that are derived from the total energy calculated by DFT, such as elastic- and lattice constants and bonding energy (BE),[12, 13] with properties obtained from experiment. The properties obtained from the DFT calculations are derivatives of the most fundamental result in the DFT framework, e.g. the electron density $n(\mathbf{r})$. Disagreements between experiment and DFT prediction could therefore stem either from lack of accuracy of the experimental results or be due to inadequacies of the functionals used. In the later case, it is difficult to pinpoint the origin at the electronic structure level, as only the theoretical properties that have been derived from the $n(\mathbf{r})$, representing more indirect metrics than the $n(\mathbf{r})$ itself, can be compared to experimentally determined metrics in validation. For instance, a disagreement in lattice constant may originate from an inaccurately calculated ground state $n(\mathbf{r})$ or from approximations made in the total energy calculation. Hence, conventionally utilized validation approaches involving the indirect comparison of properties that are calculated as derivatives of $n(\mathbf{r})$ with available experimental data or metrics fail to discriminate between these possible sources of disagreement. Additionally, agreement between DFT prediction and experiment could result from fortuitous cancellation of errors for that particular application and therefore be coincidental. A more direct approach to validation of DFT functionals that utilizes a comparison of the theoretically calculated $n(\mathbf{r})$ and experimental $n(\mathbf{r})$ would be desirable.

A possible metric is the use of structure factors, $F_g$, which can be obtained experimentally from X-ray diffraction (XRD) and are components of the Fourier transform (FT) of $n(\mathbf{r})$. Due to the lattice defects typically present in the sizable single crystal samples used in XRD experiments, determination of low order $F_g$ suffers from insufficient precision and high order $F_g$ are not sufficiently sensitive to distinguish the slight difference in the charge density obtained with different, competing density functionals.[14, 15] Quantitative convergent beam electron diffraction methods (QCBED) of transmission electron microscopy (TEM) can provide very precise and accurate low-order $F_g$, as nano-scale defect-free sample volumes can be readily used for measurements with electron beam probes as small as 0.5nm in diameter.[15] Crystalline defects that would significantly and detrimentally contribute to scattering, e. g. grain boundaries, dislocations and planar faults, can readily be identified using diffraction-contrast imaging techniques of TEM and hence can be avoided.[16] Electron structure factors can be converted into X-ray structure factors through the Mott formula.[17] QCBED methods have been successfully used for direct observations of electronic structure details, e.g. bonding related materials effects, such as d-orbital holes in $Cu_2O$[18] and the maximum difference charge density accumulation in tetrahedral rather than octahedral interstitial sites of pure Al.[19] Experimental $F_g$ measurements by QCBED have also been used to interpret bonding for many different materials, such as Cu,[20] NiO,[8] and γ-TiAl.[21] Recently, a multi-beam off-zone axis (MBOZA) QCBED method has been applied successfully for simultaneous measurement of $F_g$ and Debye Waller factors (DWFs) with unprecedented high accuracy for binary intermetallics (as high as 0.05% for some low order $F_g$).[22-24] The simultaneous determination of accurate DWFs by CBED is important for the conversion of the $F_g$, which are measured experimentally at finite temperatures, to static $F_g$ that can be compared directly to the equivalent set of $F_g$ obtained from DFT calculations. Furthermore, simultaneous determination of $F_g$ and DWFs from a single sample volume by QCBED avoids possible non-systematic error sources associated with uncertainties introduced by the investigation of two different sample volumes in separate experiments when DWFs are determined by XRD and $F_g$ by CBED. Hence, we utilize the MBOZA QCBED method for simultaneous high accuracy and precision measurements of DWFs and low-order $F_g$.[22-25] Low-order $F_g$ are introduced as facile experimental metrics for comparison with DFT calculations and validations, which can be more readily determined by experiment than alternative metrics, such as elastic constants for instance.

The MBOZA method of QCBED uses electron beam-sample orientations where several low-order reflections satisfy their respective Bragg conditions exactly,[22-24] rendering strong dynamical interactions between the incident electron beam and several excited diffracted beams. This produces CBED patterns that are very sensitive to the $F_g$ of the strongly excited reflections and the DWFs. Unlike previous QCBED approaches that use measurements from different areas of a single diffraction pattern to calculate the standard deviation,[8] with the MBOZA method it is possible to statistically evaluate $F_g$ determined from different CBED patterns for a range of experimental conditions, e.g. differing orientations and sample foil section thickness, to obtain the mean value and the standard deviation. The low-order $F_g$ refined from patterns acquired from different thicknesses and orientations are generally very consistent.[22-24]

The combination of readily available accurate acquisition of QCBED patterns through current generation TEM instrumentation and the significantly improved refinement capabilities resulting from the readily available access to powerful computers enables the systematic measurement of highly precise and accurate $F_g$ of elemental transition metals and transition metal based intermetallics in reasonable time frames. The experimentally determined $F_g$ can then be used as gauge, directly related to the electron density, to validate different DFT functionals by comparison with the equivalent calculated $F_g$. This study focused on the LDA and GGA functionals that are widely used for metals. In principle, this approach to validation of DFT methods can be extended to other sets of materials, including metal oxides or carbides with strongly correlated electron systems, and also to functionals that have been specifically designed for band gap calculations, e. g. LDA+U,[10] GGA+U and hybrid functionals.[26]

23 low order $F_g$ have been determined experimentally for five different metal elements (Cr, Fe, Co, Ni, Cu) and for three binary intermetallics (γ-TiAl, β-NiAl and $γ_1$-FePd) and will be compared with results from DFT simulations. For the prototypical BCC metals, Cr and Fe, FCC metals, Cu and Ni, and the HCP metal, Co, the interatomic bonding predominantly involves 3d-3d electron interactions. The study of γ-TiAl, β-NiAl and $γ_1$-FePd is interesting due to differences in the $3d^2$-3p, $3d^8$-3p and 3d-4d bond electron interactions in these intermetallic compounds. While γ-TiAl and $γ_1$-FePd are FCC-related tetragonal chemically ordered $L1_0$ structures, β-NiAl is a chemically ordered BCC-related B2 structure. Fe, Co, Ni and FePd are ferromagnetic. Cr is anti-ferromagnetic. The group of materials studied here represents a diverse set of different crystal structures, properties and electron interactions to explore the validation of popular DFT functionals by direct comparison of calculated and experimentally determined low-order $F_g$. The difference between experimentally determined and theoretical $F_g$ will be evaluated using the mean unsigned error (MUE).

The DFT functionals used for validation in this paper are standard LDA, LDA+U, GGA functionals including PBE,[3] WC,[27] PW91,[28] PBEsol,[11] second order GGA (SOGGA),[29] EV93,[30] and HTBS.[31] PBE and PW91 are the most popular GGA functionals used for calculating properties for crystalline solids. PBEsol, SOGGA and HTBS are GGA functionals recently designed to yield improved lattice constants and bulk moduli.[32] The LDA+U method was implemented using the self-interaction correction (SIC) approach proposed by Anisimov et al.[6] Historically the LDA+U method has been widely used for semiconductors and insulators,[6] for which the selection of U is relatively simple because it can be effectively tuned, in order to match the experimentally measured band gap. For metallic systems the selection of U is, due to the lack of a band gap, not as straightforward.[33] The use of LDA+U for metallic system has rarely been selected but would appear possible if another physical quantity can be used to aid the selection of U. For example, LDA+U has been used for calculation of the magneto-crystalline anisotropy energy for the chemically ordered tetragonal $L1_0$ phases of CoPt and FePt.[34] Inclusion of LDA+U in the comparison of DFT functionals here can potentially yield information on the selection of U for metallic systems using the $F_g$ as an experiment based metric for validation.

## II. METHODS

TEM samples for the pure metals were electro-polished from commercially available foils from Alfa Aesar. Zero-loss-energy-filtered CBED patterns have been acquired with a JEOL JEM2100F TEM instrument equipped with a post-column energy filter (Gatan Tridiem) using an energy-selecting slit of 10 to 15eV. A double-tilt low-background cold stage was used to cool the TEM specimens to temperatures as low as liquid nitrogen temperature, ~ 96K (–177 °C). An electron-beam diameter of 0.5 nm was selected in order to eliminate detrimental intensity variations that can arise in CBED patterns from thickness changes across the illuminated area. CBED patterns were recorded on a charge-coupled device (CCD) camera with a maximum resolution of 2048×2048 pixels$^2$. Discs with excitation errors *s* close or equal to zero were refined due to their advantageous signal-to-noise ratio. The intensity of each point in the experimentally obtained CBED discs is directly extracted from the image file. Each point in the selected discs is associated with a beam direction, which is used to calculate a theoretical intensity based on Bloch-wave methods.[22, 35-37] The background signal around the discs is negligible (about 0.5% of the peak intensity in energy-filtered QCBED patterns), i.e., inelastic scattering is small at the temperature of 96 K. Additionally, as peak positions are more important than the absolute intensity of peaks,[38] it is reasonable to treat the background as constant inside the discs. The program MBFIT[39] was used for the QCBED refinements. A detailed description of simultaneous refinement of DWFs and $F_g$ can be found in Ref. 24, 25. For each material, the lattice constants used in the QCBED refinements and in DFT calculations were equivalent to ensure comparability. The lattice constants at the measurement temperature, which were calculated from room temperature lattice constants[40] and expansion coefficients[41], used in this paper were: a = 2.877Å for Cr, a = 2.863Å for Fe, a=2.503Å, c=4.059Å for Co, a = 3.511Å for Ni and a = 3.610Å for Cu.

The DFT calculated theoretical $F_g$ were obtained with the WIEN2K package, which is based on an linearized augmented plane wave (LAPW) + local orbitals (lo) method.[42, 43] The $V_{xc}$ was approximated using different functionals that are available in the latest WIEN2K package. A total of 10,000 k-points in the unit cell with $R_{MT}* k_{max} = 10$ were used. These parameters have been checked for convergence. Spin-orbital coupling was not included in the calculations as its influence on $F_g$ is negligible for the materials considered here.[20]

## III. RESULTS AND DISCUSSION

## A. Thermal Vibration Amplitudes

The DWFs for Cr, Fe, Co, Ni and Cu were measured simultaneously with low order $F_g$ using the MBOZA method described in Ref. 23. DWFs were then used to subtract the thermal vibration influence on $F_g$ in order to obtain static $F_g$ for direct comparison with DFT simulation results. Thus, the accuracy of DWFs used for conversion is important to create accurate $F_g$ datasets suitable for DFT validation. For the isometric cubic crystal structures,[44] such as Cr, Fe, Ni and Cu, a single DWF suffices, while the uniaxial hexagonal Co requires the measurement of two anisotropic DWFs, $B_{11}$ and $B_{33}$. The current experimental DWFs are compared with previous measurements in Table I. The XRD values were converted from room temperature values[45] to the temperatures at which the CBED patterns were acquired using a Debye model.[44] For Fe and Ni, the XRD DWFs agree well with CBED DWFs. For Cr, the CBED DWF is much larger than the XRD DWF. For Cu, the CBED DWF is slightly smaller than the XRD DWF. There is no general systematic trend that can explain the differences between XRD and CBED measurements. A possible source for the technique dependent differences in DWF values would be significant differences in the defect content of the sampled crystal volume. XRD measurements require large single crystalline samples with commonly not very well quantified defect concentrations, which therefore are difficult to treat during refinements, whereas in CBED measurements defect-free nano-scale volumes are sampled. Therefore, simultaneous measurement of the DWF from a perfect single crystal sample area at that specific temperature results in improved CBED $F_g$ determination. To the best of our knowledge, the DWFs for Co have not been reported, suggesting that even for less commonly studied materials, quantitative CBED experiments can be an efficient and appropriate method for DWF and low order $F_g$ determination.

## B. Experimental low order $F_g$

For the cubic structures low order $F_g$ were measured from patterns acquired close to a <110> zone axis orientation, which contain all the important low order $F_g$ ($F_{g110}$, $F_{g200}$ and $F_{g211}$ for BCC; $F_{g111}$, $F_{g200}$ and $F_{g220}$ for FCC) using the MBOZA QCBED method. In addition, $F_g$ were also measured from patterns along other orientations (<100>, <111> et. al) to evaluate the accuracy of the MBOZA method.

Fig. 1 shows an example of a QCBED refinement result for a pattern acquired close to the [110] zone axis for FCC Ni. The difference between theory and experiment is measured using the R value.[39] A perfect match is achieved when the R value is equal to 0. Two refinements were performed consecutively. First, $F_g$ were set to values based on independent atom model (IAM)[46] and only the DWF's were relaxed (column II). The R value is 0.17. In the second refinement, both the DWF and $F_g$ ($F_{g111}$, $F_{g200}$ and $F_{g220}$) were relaxed (column III). The R value decreases from 0.17 to 0.14, which is also represented in the difference maps shown in Fig. 1. Column III-I shows weaker and reduced contrast than column III-I, reflecting the improved fit. The decrease of the R value indicates that including the $F_g$ can improve the match between theory and experiment. The improvement of R value for the elemental metal Ni is much less significant than for the intermetallics (for example, R decreases from 0.36 to 0.14 for β-NiAl upon the second refinement),[23] suggesting that the change of $F_g$ caused by bonding in the elemental metal Ni is very small as compared to the case of the intermetallics. This implies that the measurement of $F_g$ requires very high accuracy and precision for elemental metals. For each disc, intensities of 195303 pixels are used, which is a sufficiently large dataset for robust refinements.

Static low order $F_g$ for the five materials considered here are summarized in Table II. Our measurement results for metals are reported here for the first time. The mean values and the standard deviations for each $F_g$ listed in the table have been calculated from refinement results of 10-20 patterns from different thicknesses of the TEM foil sections probed by CBED. The accuracy of our measurement is generally on the order of 0.1%, which is unprecedented and sufficient to distinguish between different DFT functionals. Using low order $F_g$ ($h^2+k^2+l^2 \leq 4$) for validation of DFT functionals is very important. In order to develop a functional that can yield an electron charge density that can correctly describe both the structure and properties of a material, it is essential that this electron charge density, which is derived from that functional, can predict and describe low order $F_g$ most accurately, as low order $F_g$ are strongly influenced by bonding. The bonding is an essential and important feature of a crystal structure. Therefore, the ability to predict correctly the low order $F_g$ would appear to be a necessary condition, but probably not a sufficient condition (as this electron charge density might be able to predict low order $F_g$ but no other properties), for the robust description of the structure of a material. Also, the accuracy of the CBED methods for low order $F_g$ is inherently superior to that of high order $F_g$ due to a scale factor proportional to $\mathbf{g}^2$ in the Mott formula, which converts measured electron diffraction $F_g$ to XRD $F_g$.[17] Therefore, although high order $F_g$ could potentially contribute to understanding of effects originating from core electrons, experimental measurements by CBED do not have sufficient accuracy for validation of different DFT functionals and thus are not included in Table II.

### C. Theoretical low order $F_g$

The DFT based theoretical $F_g$, obtained from WIEN2K package calculations with different functionals, are listed in Table III. Here we extend the comparison to include three intermetallics β-NiAl, γ-TiAl and $γ_1$-FePd. Experimental $F_g$ data for the three intermetallics has been reported previously.[23-25] The IAM $F_g$ based on the Doyle and Turner method[46] are listed for comparison. $F_g$ from γ-ray diffraction for Cr,[47] Fe,[48] Co,[49] and Ni[50] are included in Table III. LDA+U has only been used for Ni, Cu, β-NiAl and $γ_1$-FePd as LDA already performs well enough for the other materials. For the LDA+U calculations the U values for the different materials were adjusted until the low order $F_g$ obtained by conversion from the calculated electron charge density yield values closest to the respective experimentally measured CBED $F_g$.

The MUEs for the $F_g$ calculated for each of the different DFT functionals compared with the experimentally measured CBED $F_g$ are shown in Table IV. The average uncertainty of the experiment is calculated and listed in the column CBED MUE. MUE values less than the CBED MUEs are considered as good agreement between $F_g$ results obtained from DFT calculations and CBED experiments. The overall MUE for the IAM is six times larger than the CBED and DFT MUE, indicating that there is a bond effect on the structure factors, which causes deviations of $F_g$ from IAM values that both CBED refinements and DFT calculations capture. The MUE for the intermetallics is larger than the MUE for metals, suggesting stronger effects from bonding for the intermetallics. The γ-ray results do not agree with either CBED or DFT results. This might be due to the detrimental effects on the $F_g$ measurements from significant populations of imperfections expected to be contained in the single crystals samples used for the γ-ray diffraction experiments, which were cubes of dimensions approximately 2.5mm×2.5mm×2.5mm.[47-50]

LDA and all the GGA functionals, WC, PBE, PW91 and HTBS perform reasonably well in the prediction of a ground state electron density (MUE less than 0.06 with an experimental uncertainty of 0.03) that can predict the experimentally determined low order $F_g$. PW91 is slightly better than the others with MUE=0.055. The overall performance of PBEsol, SOGGA and EV93 is also reasonably good. For metals, PW91 and EV93 have the best performance with MUE=0.032. For intermetallics, PW91 has the best performance among all the GGA functionals. Although CBED MUE only slightly increases from 0.027 for metals to 0.033 for intermetallics, the performance for all the GGA functionals is significantly better for metals (PW91 MUE=0.032) than for intermetallics (PW91 MUE=0.075).

For the metals with different crystal structures, different GGA functionals perform better for certain structures. For the BCC metals Cr and Fe, LDA, WC, PBE, PW91 and HTBS agree well with our experiment. SOGGA, EV93 results deviate significantly from CBED. For the HCP metal Co all the functionals agree very well with our measurements. For the FCC metals Ni and Cu, EV93 performs the best, while the other functionals have MUE two times larger than the CBED MUE. The disagreement between GGA and CBED $F_g$ was also reported for Cu by Saunders et al..[51] For β-NiAl and $γ_1$-FePd, LDA and all the GGA functionals yield very large MUE compared with experimental uncertainty. For γ-TiAl all the functionals provide similar MUEs and all of them are comparable to the experimental uncertainty. For $γ_1$-FePd, all the GGA functionals predict $F_g$ significantly different from the CBED $F_g$.

The EV93 functional was designed for accurate $V_x$ calculations, which is considerably better than GGA, and the $V_c$ is from standard LDA.[52] In comparison to all other DFT functionals EV93 was reported to agree best with experimental $F_g$ for FCC Si[53] and HCP Mg with a close packed structure.[54] The energy part of EV93 based calculations shows poor performance.[52] EV93 significantly overestimates the lattice constants for transition metals Fe and Nb.[52] However, it has been reported that EV93 can improve band structure calculations, which mainly depends on the $V_x$. A good $V_x$ might be the reason why EV93, in comparison to other potentials, predicts significantly better the $F_g$ for Ni and Cu.

Notably, results for $F_g$ obtained by LDA+U (SIC) are superior to all the GGA functionals with an overall ("Total" row in Table IV) MUE=0.034. For Cr, Fe, Co and TiAl, LDA is sufficient as it agrees well with CBED $F_g$. For Ni, Cu and β-NiAl, the orbital potential U cannot be neglected due to almost fully filled 3d orbitals. For $γ_1$-FePd, the free atom configurations of Fe has six 3d electrons and that of Pd has a fully filled 4d orbital, indicating that a strong orbital potential contribution should be included for this system. The big improvement of LDA+U over LDA and all the GGA functionals shows that the consideration of the orbital potential U can, at least as far as the ability to predict low order $F_g$ is concerned, not only lead to currently superior results for semiconductors but also for some metal and intermetallic systems. This would be consistent with prior studies reporting for HCP Mg that the core electron density agrees better with experiment when SIC is considered.[54] Also, Blaha et al. found that the electric field gradients of cuprates in the metallic state can be better predicted by DFT using a U term.[55] The excellent performance of LDA+U reported here also indicates that the experimental metric of CBED $F_g$ can be efficiently used for conducting metallic materials to provide starting values, otherwise difficult to obtain, and assist effectively in tuning the U term for electronic structure calculations of these materials.

None of the GGA functionals agree with the CBED $F_g$ measurements for $γ_1$-FePd. The GGA functionals especially fail in predicting the superlattice reflection (SR) $F_g$, namely $F_{g001}$ and $F_{g110}$. The difference between the actual $F_g$ for a crystal and the IAM based $F_g$ is usually interpreted as bond related. GGA calculation based $F_{g001}$ and $F_{g110}$ deviate more from the IAM values than the experimental CBED $F_g$ values, which indicates that the GGA overestimates the bonding effects. This observation is not surprising, as in GGA the electron self − interaction with its own charge density is generally considered to result in an overestimation of delocalization,[13] which raises the energy for localized states, and has been illustrated for example in the case of TiAl in electron charge density deformation maps.[25] The amplitudes of CBED SR $F_g$ are generally larger than the amplitudes of GGA SR $F_g$ for the chemically ordered intermetallics γ-TiAl, β-NiAl and $γ_1$-FePd. A long range order parameter reduced to less than one can only decrease the amplitude of the SR $F_g$. Hence, the argument that an inexact order parameter used in the experiments is responsible for the discrepancy between DFT and CBED measurements of the low order SR $F_g$ cannot be applied here.

LDA+U is the only DFT approach that accurately predicts SR $F_g$, namely $F_{g001}$ and $F_{g110}$, for $\gamma_1$-FePd, indicating that potentially LDA+U could be used for significantly improved DFT calculations of this intermetallic systems as compared to the more conventional GGA based DFT. This would be consistent with a previous report on use of a +U approach to DFT calculations for $\gamma_2$-FePt [34]. The tuning of the U term can be assisted by matching to CBED $F_g$ as demonstrated here. Additional work is in progress to evaluate the prediction of materials properties using the LDA+U functional obtained from the initial CBED $F_g$ based optimization.

## IV. SUMMARY

It is a necessary but not a sufficient condition that the electron charge density and its corresponding ground state energy, which are supposed to describe the structure and the properties of a material correctly, have to be able to predict low order structure factors correctly. Therefore, the direct comparison of sufficiently accurate and precise experimentally determined low order $F_g$ with corresponding $F_g$ calculated by DFT provides an excellent metric for initial validation. Here, low order $F_g$ for five pure metal elements with three different crystal structures and three binary intermetallics with two different crystal structure were measured with unprecedented accuracy and precision using the MBOZA QCBED method, providing a large dataset for validation of the most fundamental property, e.g. the electron density, by direct comparison with calculations performed with different DFT functionals. The CBED and DFT results generally agree very well with each other for most materials, which indicates that both DFT and CBED can reach sufficient accuracy to reveal the electron density for many-electron systems. Out of all the GGA functionals, PW91 has the best overall performance, while EV93 performs the best for the FCC metals Ni and Cu. The LDA+U approach is superior to all the GGA functionals that we tested. For the pure metals Ni and Cu, and the intermetallics $\beta$-NiAl and $\gamma_1$-FePd, the d-orbital potential cannot be neglected. We introduced a promising approach to select the U term for metal-like conducting systems, e. g. to match experimental $F_g$, analogous to matching the band gap for semiconductors and insulators. The easy access to directly comparable highly accurate and precise experimental $F_g$ can assist and inspire the development of improved DFT functionals.

## ACKNOWLEDGEMENTS

This work is supported by a grant of the Office of Basic Energy Sciences, division of Materials Science and Engineering, of U.S. DOE (Grant No DE-FG02-08ER46545). This work was also supported by the Center for Molecular and Materials Simulations at the University of Pittsburgh.

Table I. DWFs for transition metal elements at 100K (the unit is Å$^2$)

| DWFs | B(Cr) | B(Fe) | $B_{11}$(Co) | $B_{33}$(Co) | B(Ni) | B(Cu) |
|---|---|---|---|---|---|---|
| MBOZA CBED | 0.17(2) | 0.150(9) | 0.18(1) | 0.14(1) | 0.17(1) | 0.51(2)[a] |
| XRD [45b] | 0.130 | 0.160 | | | 0.165 | 0.567 |

[a] CBED patterns were acquired at 300K for Cu and 100K for the other metals.

[b]Values at certain temperatures are calculated using Debye model and Debye temperature in the literature.

Figure 1 An experimental CBED pattern acquired from FCC Ni near [110] zone axis (a). The thickness of the area was refined to be 150.9nm. (b) The difference from using IAM $F_g$ in the simulation is shown in (III-I) and the improved agreement after relaxing $F_g$ in the refinement is shown in (II-I).

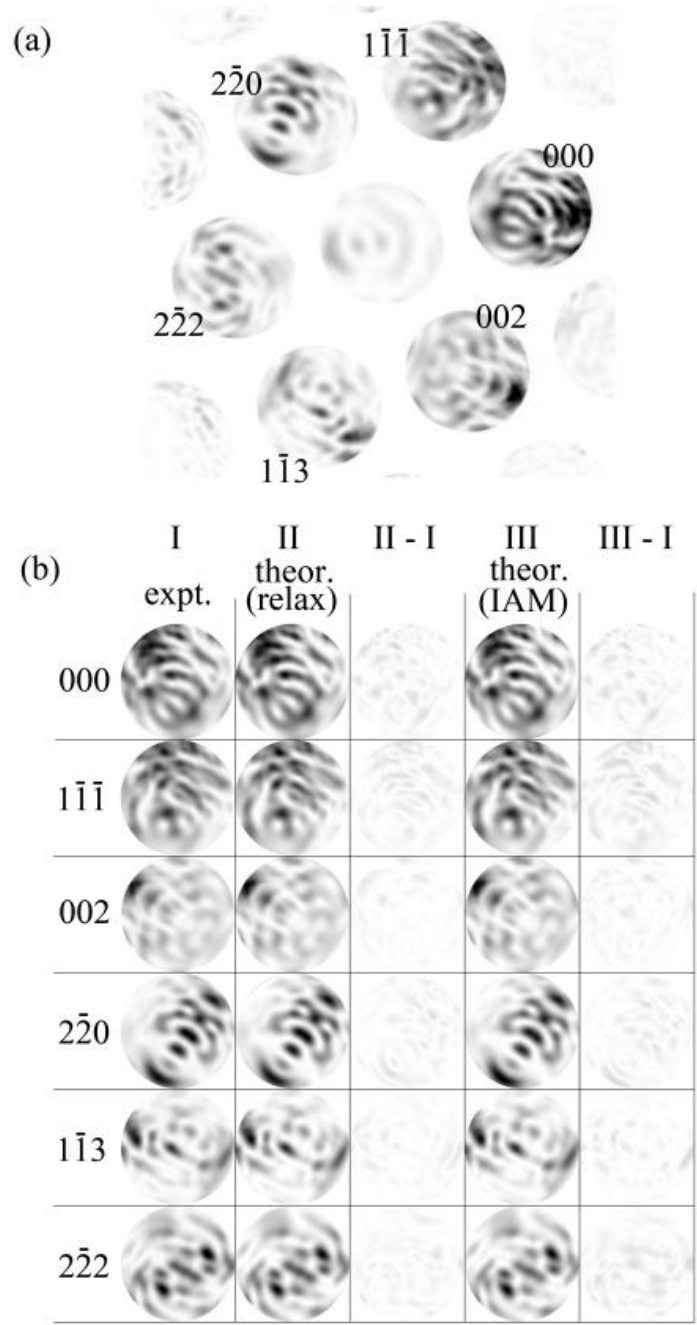

Table II. Experimental $F_g$ measured using MBOZA CBED method. $F_g$ from <100>, <110> and <111> zone axes were measured for cubic structures. For hexagonal Co, $F_g$ from <001>, was also measured.

| Metals | $F_g$(hkl) | <100> | <110> | <111> | <001> |
|---|---|---|---|---|---|
| Cr | $F_{g110}$ | | 16.33(2) | | |
| | $F_{g200}$ | | 13.39(4) | | |
| Fe | $F_{g110}$ | | 18.311(8) | 18.30(2) | |
| | $F_{g200}$ | | 15.17(3) | | |
| Co | $F_{g100}$ | -9.98(4) | | | -9.97(2) |
| | $F_{g002}$ | -19.41(5) | | | |
| | $F_{g101}$ | -16.29(2) | | | |
| Ni | $F_{g111}$ | | 20.46(2) | | |
| | $F_{g200}$ | | 19.13(2) | | |
| Cu | $F_{g111}$ | | 21.80(3) | | |
| | $F_{g200}$ | 20.43(3) | 20.47(4) | | |

Table III. Compilation of experimental and theoretical $F_g$ for metals and intermetallics

|  | $F_g(hkl)$ | CBED | IAM | γ-ray | LDA | WC[a] | PBE | PW91 | PBEsol | SOGGA[b] | EV93[c] | HTBS[d] | LDA+U[e] |
|---|---|---|---|---|---|---|---|---|---|---|---|---|---|
| Cr | $F_{g110}$ | 16.33(2) | 16.705 | 17.093 | 16.320 | 16.323 | 16.327 | 16.335 | 16.316 | 16.307 | 16.363 | 16.325 | |
|  | $F_{g200}$ | 13.39(4) | 13.574 | 13.871 | 13.429 | 13.438 | 13.445 | 13.446 | 13.434 | 13.434 | 13.474 | 13.440 | |
| Fe | $F_{g110}$ | 18.311(8) | 18.462 | 19.016 | 18.309 | 18.308 | 18.312 | 18.320 | 18.302 | 18.293 | 18.372 | 18.309 | |
|  | $F_{g200}$ | 15.17(3) | 15.250 | 15.614 | 15.145 | 15.149 | 15.155 | 15.156 | 15.147 | 15.148 | 15.205 | 15.152 | |
| Co | $F_{g100}$ | -9.97(2) | -10.015 | -10.073 | -9.969 | -9.968 | -9.970 | -9.975 | -9.965 | -9.959 | -9.990 | -9.969 | |
|  | $F_{g002}$ | -19.41(5) | -19.498 | -19.483 | -19.364 | -19.363 | -19.367 | -19.376 | -19.357 | -19.347 | -19.409 | -19.364 | |
|  | $F_{g101}$ | -16.29(2) | -16.443 | -16.509 | -16.309 | -16.309 | -16.313 | -16.319 | -16.304 | -16.297 | -16.348 | -16.310 | |
| Ni | $F_{g111}$ | 20.46(2) | 20.513 | 20.619 | 20.408 | 20.408 | 20.413 | 20.421 | 20.402 | 20.393 | 20.458 | 20.412 | 20.469 |
|  | $F_{g200}$ | 19.13(2) | 19.214 | 19.349 | 19.076 | 19.082 | 19.088 | 19.093 | 19.077 | 19.072 | 19.133 | 19.085 | 19.129 |
| Cu | $F_{g111}$ | 21.80(3) | 22.065 | | 21.710 | 21.712 | 21.717 | 21.724 | 21.707 | 21.699 | 21.761 | 21.713 | 21.794 |
|  | $F_{g200}$ | 20.45(4) | 20.710 | | 20.382 | 20.388 | 20.395 | 20.399 | 20.384 | 20.380 | 20.440 | 20.390 | 20.450 |
| β-NiAl | $F_{g100}$ | 13.722(8) | 13.501 | | 13.664 | 13.662 | 13.658 | 13.666 | 13.659 | 13.650 | 13.674 | 13.676 | 13.733 |
|  | $F_{g110}$ | 28.98(2) | 29.105 | | 28.923 | 28.923 | 28.925 | 28.934 | 28.915 | 28.905 | 28.953 | 28.919 | 28.974 |
| γ-TiAl[f] | $F_{g001}$ | 8.017(7) | 7.916 | | 8.018 | 8.015 | 8.008 | 8.013 | 8.012 | 8.008 | 7.999 | 8.028 | |
|  | $F_{g110}$ | 7.16(4) | 7.223 | | 7.077 | 7.074 | 7.070 | 7.073 | 7.072 | 7.069 | 7.074 | 7.085 | |
|  | $F_{g111}$ | 24.30(1) | 24.524 | | 24.289 | 24.282 | 24.281 | 24.295 | 24.272 | 24.257 | 24.299 | 24.274 | |
|  | $F_{g002}$ | 22.99(4) | 23.245 | | 23.018 | 23.023 | 23.027 | 23.033 | 23.016 | 23.010 | 23.056 | 23.022 | |
|  | $F_{g200}$ | 22.67(1) | 23.031 | | 22.714 | 22.715 | 22.717 | 22.721 | 22.708 | 22.704 | 22.734 | 22.712 | |
| γ-FePd | $F_{g001}$ | -18.60(2) | -18.883 | | -18.378 | -18.387 | -18.393 | -18.391 | -18.390 | -18.391 | -18.415 | -18.392 | -18.592 |
|  | $F_{g110}$ | -17.50(6) | -17.682 | | -17.296 | -17.309 | -17.318 | -17.316 | -17.311 | -17.312 | -17.354 | -17.314 | -17.499 |
|  | $F_{g111}$ | 54.32(4) | 54.702 | | 54.252 | 54.255 | 54.269 | 54.284 | 54.244 | 54.224 | 54.384 | 54.256 | 54.370 |
|  | $F_{g200}$ | 51.37(6) | 51.760 | | 51.355 | 51.361 | 51.375 | 51.382 | 51.353 | 51.343 | 51.469 | 51.361 | 51.354 |
|  | $F_{g002}$ | 50.46(8) | 50.922 | | 50.585 | 50.600 | 50.618 | 50.625 | 50.591 | 50.582 | 50.727 | 50.603 | 50.853 |

[a]GGA WC for exchange and GGA PBE for correlation

[b]GGA SOGGA for exchange and GGA PBE for correlation

[c]GGA EV93 for exchange and LDA for correlation

[d]GGA HTBS for exchange and GGA PBE for correlation

[e]U=0.5Ry for metal Ni and Cu. U(Ni)=0.3Ry in β-NiAl. U(Fe)=U(Pd)=0.5Ry in γ-FePd.

[f]A tp4 cell is used for tetragonal $L1_0$ TiAl and FePd.

Table IV. Difference between CBED dataset and previous experimental and current theoretical datasets

| MUE | CBED[a] | IAM | γ-ray | LDA | WC | PBE | PW91 | PBEsol | SOGGA | EV93 | HTBS | LDA+U[d] |
|---|---|---|---|---|---|---|---|---|---|---|---|---|
| Cr | 0.030 | 0.280 | 0.622 | 0.024 | 0.027 | 0.029 | 0.031 | 0.029 | 0.034 | 0.059 | 0.028 | 0.024 |
| Fe | 0.019 | 0.116 | 0.575 | 0.014 | 0.012 | 0.008 | 0.011 | 0.016 | 0.020 | 0.048 | 0.010 | 0.014 |
| Co | 0.030 | 0.095 | 0.132 | 0.022 | 0.022 | 0.022 | 0.023 | 0.024 | 0.027 | 0.027 | 0.023 | 0.022 |
| Ni | 0.020 | 0.068 | 0.189 | 0.053 | 0.050 | 0.044 | 0.038 | 0.055 | 0.063 | 0.003 | 0.047 | 0.005 |
| Cu | 0.035 | 0.262 |  | 0.079 | 0.075 | 0.069 | 0.064 | 0.079 | 0.086 | 0.025 | 0.074 | 0.003 |
| β-NiAl | 0.014 | 0.173 |  | 0.057 | 0.059 | 0.060 | 0.051 | 0.064 | 0.073 | 0.038 | 0.053 | 0.009 |
| γ-TiAl | 0.021 | 0.201 |  | 0.034 | 0.037 | 0.040 | 0.038 | 0.037 | 0.039 | 0.047 | 0.037 | 0.034 |
| γ-FePd | 0.052 | 0.340 |  | 0.127 | 0.124 | 0.121 | 0.122 | 0.125 | 0.128 | 0.152 | 0.122 | 0.094 |
| M[b] | 0.027 | 0.158 | 0.352 | 0.037 | 0.036 | 0.033 | 0.032 | 0.039 | 0.044 | 0.032 | 0.035 | 0.014 |
| IM[c] | 0.033 | 0.254 |  | 0.076 | 0.077 | 0.077 | 0.075 | 0.078 | 0.082 | 0.089 | 0.075 | 0.054 |
| Total | 0.030 | 0.208 | 0.352 | 0.057 | 0.057 | 0.056 | 0.055 | 0.060 | 0.064 | 0.062 | 0.056 | 0.035 |

[a]MUE=$\frac{1}{N}\sum_{i=1}^{N}\sigma_i$. For the other columns, MUE=$\frac{1}{N}\sum_{i=1}^{N}\left|F_i^{CBED}-F_i^{theor.}\right|$

[b]MUE for five metals

[c]MUE for three intermetallics

[d]LDA results were used to calculate MUE for materials that LDA+U was not performed on.